\documentclass{PoS}

\usepackage{amsmath, amsthm, amssymb}
\usepackage[T1]{fontenc}

\newcommand{\erfc}{\text{erfc}}

\newcommand{\fig}[1]{Fig.~#1}

\renewcommand{\l}{\left}
\renewcommand{\r}{\right}
\renewcommand{\d}{\text{d}}

\newcommand{\rmax}{r_{\max}}

\newcommand{\legosphere}{LEGO$^{\text{\textregistered}}$ sphere}

\newcommand{\vk}{\textbf{k}}
\newcommand{\vr}{\textbf{r}}
\newcommand{\vn}{\textbf{n}}
\renewcommand{\v}[1]{\textbf{#1}}

\title{Application of Ewald's Method for Efficient Summation of Dyon Long-Range Potentials}

\ShortTitle{Application of Ewald's Method for Efficient Summation of Dyon Long-Range Potentials}

\author{\speaker{Benjamin Maier}, Michael M\"uller-Preussker \\
        Humboldt-Universit\"at zu Berlin, Institut f\"ur Physik, Newtonstr. 15, D-12489 Berlin, Germany}

\author{Falk Bruckmann, \\
        Universit\"at Regensburg, Institut f\"ur Theoretische Physik, D-93040 Regensburg, Germany}

\author{Simon Dinter, \\
        NIC, DESY Zeuthen, Platanenallee 6, D-15738 Zeuthen, Germany}

\author{Ernst-Michael Ilgenfritz, \\
        Joint Institute for Nuclear Research, VBLHEP, 141980 Dubna, Russia}

\author{Marc Wagner, \\
        Goethe-Universit\"at Frankfurt am Main, Institut f\"ur Theoretische Physik, \\ $\quad$ Max-von-Laue-Stra{\ss}e 1, D-60438 Frankfurt am Main, Germany}

\abstract{We study a model of dyons for SU(2) Yang-Mills theory at finite temperature $T<T_c$, in particular its ability to generate a confining force between a static quark antiquark pair. The interaction between dyons corresponds to a long-range $1/r$ potential, which in naive treatments with a finite number of dyons typically gives rise to severe finite volume effects. To avoid such effects we apply the so-called Ewald method, which has its origin in solid state physics. The basic idea of Ewald's method is to consider a finite number of dyons inside a finite cubic volume and enforce periodicity of this volume. We explain the technicalities of Ewald's method and outline how the method can be applied to a wider class of $1 / r^p$ long-range potentials.}

\FullConference{Xth Quark Confinement and the Hadron Spectrum,\\
		October 8-12, 2012\\
		TUM Campus Garching, Munich, Germany}

\begin{document}


\section{Introduction}

We consider pure SU(2) Yang Mills theory at finite temperature in the confinement phase $T<T_c$, which is a crude approximation of Quantum Chromodynamics (QCD). We study a model of dyons, in particular the quark antiquark free energy, with the aim to better understand the phenomenon of confinement. In this work we mainly focus on technical difficulties associated with the long-range nature of the dyon potentials. Physical aspects and conclusions are discussed in detail in \cite{Bruckmann:2009nw,Bruckmann:2011yd} and another talk given at this conference \cite{MMP}.


\section{\label{SEC001}The non-interacting dyon model}

In Yang Mills theory observables $O$ are given by the path integral
\begin{equation}
\label{EQN002} \l<O\r> = \frac1Z \int\mathcal DA\ O[A]\exp\l(-S_{\text{\textrm{YM}}}[A]\r) .
\end{equation}
Since there are currently no methods to solve this path integral for low-energy observables analytically, one either resorts to numerical lattice gauge theory or to certain simplifying approximations.

One such approach particularly useful to obtain a qualitative understanding of certain phenomena of Yang-Mills theory and QCD is the semi-classical approximation. One expands the path integral around classical solutions of the Yang-Mills field equations, for which the action is locally minimized, and which are expected to dominate the path integral. A specific kind of a semi-classical model and its capability to generate confinement is based on dyons. Dyons are localized objects carrying electric charge as well as magnetic charge and are named after particles with similar properties \cite{Schwinger:1969ib}. The path integration (\ref{EQN002}) is transformed from field coordinates to dyon collective coordinates and quantum fluctuations, where a Jacobian emerges. Part of this Jacobian is the determinant of the so-called moduli space metric. This moduli space metric has been calculated analytically for calorons \cite{Kraan:1998pn}, following \cite{KraanVanBaal1,KraanVanBaal2,LeeLu}, which are pairs of different-kind dyons. A proposal for a metric of an arbitrary number of same-kind and different kind dyons was made in \cite{Diakonov:2007nv}. Numerical investigations of this generalized metric \cite{Bruckmann:2009nw}, however, indicated certain shortcomings, in particular its non-positive-definiteness, casting severe doubts on its usefulness. Our main interest in this work is to test a numerical method able to treat the dyon long-range potentials in a proper way. Therefore, we study a much simpler model of dyons without any interactions, i.e.\ where the moduli space metric is ignored.

The key observable we are studying is the free energy of a static quark antiquark pair at separation $d = \l|\v r-\v r'\r|$, which is given by
\begin{equation}
F_{Q\bar Q}(d) = -T\ \log\l< P(\v r)P^{\dagger}(\v r')\r>.
\end{equation}
The Polyakov loop correlator $\l< P(\v r)P^{\dagger}(\v r')\r>$ is obtained as a statistical average in dyon ensembles characterized by the spatial dyon density $\rho$ and the temperature $T$ (we always consider maximally non-trivial holonomy, which seems to be intimately connected to the confinement phase \cite{Gerhold:2006sk}). The dyon ensembles we study are neutral, i.e.\ there is an identical number of dyons and antidyons. Individual dyon configurations of these ensembles are given by the randomly and uniformly chosen dyon positions $\l\{\v r_j\r\}$. One can show that in such ensembles a Polyakov loop is given by
\begin{equation}
     P(\v r) = -\sin\l(\frac1{2T}\Phi(\v r)\r),
\end{equation}
where $\Phi$ is the superposition of the $0$-component of the dyon gauge field in the Abelian limit
\begin{equation}
\label{EQN001}    \Phi(\v r) = \sum_j\frac{q_j}{|\v r_j - \v r |} .
\end{equation}
For a more detailed discussion of these equations and the non-interacting dyon model in general we refer to \cite{Bruckmann:2009nw,Bruckmann:2011yd}.

In the following sections we are mainly concerned with evaluating (\ref{EQN001}) numerically, which contains an infinite sum $\sum_j$ over $1 / r$ long-range potentials.


\section{Long-range dyon potentials and finite volume effects}

The $1/r$ long-range nature of the dyon potential causes severe problems for numerical simulations. One expects that rather large volumes are needed, to render finite volume effects negligible. This in turn amounts to a huge number of dyons, which is proportional to the required computational resources.

\begin{figure}[htb]
 \centering
 \begin{minipage}{.47\textwidth}
    \centering
    \includegraphics[width=5cm]{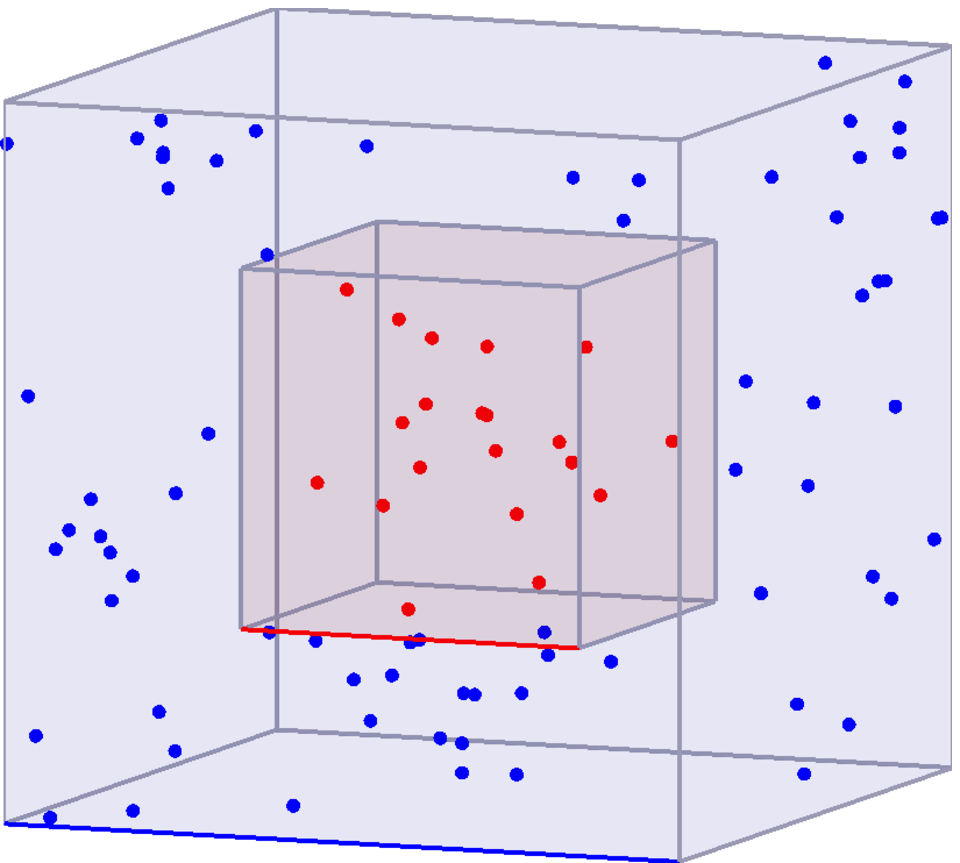}\\
    \begin{flushleft}
        \vspace{-6cm}\textbf{a)}\vspace{4.5cm}
    \end{flushleft}
    \vspace{-1.45cm}
    \Large\textcolor{red}{$\ell\ \ \ \ $}\\
    \vspace{0.4cm}
    \hspace{-1.5cm}\textcolor{blue}{$L$}\\
 \end{minipage} \hspace{\fill}
 \begin{minipage}{.47\textwidth}
    \centering
    \includegraphics[width=7cm]{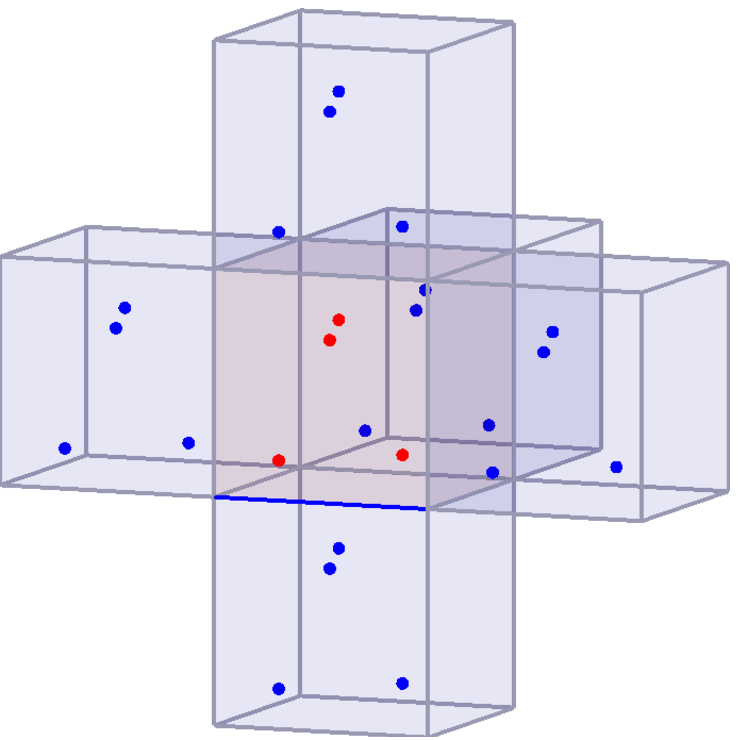}\\
    \begin{flushleft}
        \vspace{-7.3cm}\textbf{b)}\vspace{6cm}
    \end{flushleft}
    \vspace{-2.5cm}
    \hspace{-1cm}\Large\textcolor{blue}{$L$}
    \vspace{1.8cm}
 \end{minipage}    
 \caption{\label{fig:box_copies}\textbf{a)} Dyons in a cubic volume of length $L$ (blue), evaluating observables in a cubic volume of length $\ell < L$ (red). \textbf{b)} Dyons in a volume of length $L$ with periodic boundary conditions.}
\end{figure}

A first attempt to simulate dyon ensembles numerically, which suffers from the just mentioned problem, is described in \cite{Bruckmann:2009nw}. There $n_D$ dyons are considered in a cubic spatial volume of length $L$. Observables are then evaluated in a cubic spatial volume of length $\ell < L$ located at the center of the larger volume (cf.\ \fig{\ref{fig:box_copies}a}). This straightforward method, however, has certain shortcomings: (A) reducing finite volume effects to a moderate level requires $\ell \ll L$, which drastically reduces the volume, in which observables can be evaluated; this clearly increases statistical errors; (B) an extrapolation to infinite volume is technically difficult, since it has to be done with respect to two parameters ($\ell$ and $L$); (C) when attractive and repulsive forces between dyons are taken into account, dyons tend to accumulate near the boundary of the large volume, i.e.\ translational invariance is broken severely.

A better method for treating long range dyon ensembles, which solves or eases the problems mentioned above, seems to mimic infinite volume by implementing periodic boundary conditions. One considers a cubic spatial volume of length $L$ filled with $n_D$ dyons (dyon density $\rho=n_D/L^3$). This volume is periodically repeated in all three spatial directions (cf. \fig{\ref{fig:box_copies}b}). With this setting one expects (A) that finite volume effects are significantly reduced, since the original volume is ``surrounded by infinitely many dyons'' in every spatial direction; (B) an extrapolation to infinite volume only has to be done with respect to one parameter, $L$; (C) interacting dyons will not accumulate near the boundary of the volume, because periodicity implies exact translational invariance.


\section{Ewald's method}

To implement periodic boundary conditions for our long-range dyon potentials we resort to a method proposed first in the context of condensed matter physics \cite{Ewald:1921} (``Ewald's method'') and is widely used in plasma physics, as well.

The central quantity, when one is e.g.\ interested in Polyakov loop averages or the quark antiquark free energy, is the potential of $n_D$ dyons in a periodic cubic volume of length $L$,
\begin{equation}
\Phi(\vr)=\sum_{\vn\in\mathbb Z^3}\sum_{j=1}^{n_D}\frac{q_j}{\l|\vr-\vr_j+\vn L\r|}
\end{equation}
(cf.\ also section~\ref{SEC001}). Due to the long-range nature of the individual dyon potentials, $\sum_{\vn\in\mathbb Z^3} \ldots$ converges far too slowly for any straightforward efficient numerical evaluation. Ewald's method solves this problem by splitting the sum into a short-range part and a long-range part,
\begin{align}
&\Phi(\vr)=\Phi^{\text{Short}}(\vr) + \Phi^{\text{Long}}(\v r) \\
&\Phi^{\text{Short}}(\vr)=\sum_{\vn\in\mathbb Z^3}\sum_{j=1}^{n_D}\frac{q_j}{\l|\vr-\vr_j+\vn L\r|}\ \erfc\l(\frac{\l|\vr-\vr_j+\vn L\r|}{\sqrt{2}\lambda}\r)\\
&\Phi^{\text{Long}}(\v r)=\frac{4\pi}{V}\sum_{\vk\neq0}\sum_{j=1}^{n_D}\frac{q_j}{k^2}\ e^{i\vk(\vr-\vr_j)}\ e^{-\lambda^2k^2/2} .
\end{align}
For the long-range part the sum is over momenta $\v k=(2\pi/L) \v m$ with $\v m\in\mathbb Z^3$. The arbitrary parameter $\lambda$ controls the trade-off between the short-range and the long-range part. Both parts converge exponentially fast, due to $\erfc(\l|\vr-\vr_j+\vn L\r| / \sqrt{2}\lambda)$ and $e^{-\lambda^2k^2/2}$, respectively. A detailed and pedagogical derivation of the splitting into short-range and a long-range part can be found in \cite{leewei}.


\subsection{Evaluating the short-range part}

To determine Polyakov loop averages, one typically computes the dyon potential at $M\propto V$ points $\v r$ distributed throughout the volume. Due to the exponential suppression by \\ $\erfc(\l|\vr-\vr_j+\vn L\r| / \sqrt{2}\lambda)$, it is sufficient to consider dyons inside a sphere with radius $r_{\max} \propto \lambda$ and center $\v r$, i.e.\
\begin{equation}
\Phi^{\text{Short}}(\vr)=\sum_{\vn, j, |\vr-\vr_j+\vn L| < r_{\max}}\frac{q_j}{\l|\vr-\vr_j+\vn L\r|}\ \erfc\l(\frac{\l|\vr-\vr_j+\vn L\r|}{\sqrt{2}\lambda}\r) .
\end{equation}
Consequently, the total computational cost is $\mathcal O(V\lambda^3)$.

In practice $\lambda^3 \ll V$ is chosen. For an algorithm scaling according to $\mathcal O(V\lambda^3)$ one clearly needs to determine all dyons close to a point $\v r$ without iterating over all $n_D \propto V$ dyons. To this end, one divides the original volume into small cubic subvolumes. Technically, the volume is a list of subvolumes, where each subvolume corresponds to another individual list of those dyons located inside. For any point $\v r$ one can then easily determine all subvolumes inside or at the boundary of the corresponding sphere with radius $r_{\max}$ and then iterate over the relevant $\mathcal O(\lambda^3)$ dyons. The approximate sphere of subvolume copies (``\legosphere'') is shown in \fig{\ref{fig:legosphere}}. Note that the radius of the \legosphere{} is larger than $\rmax$, to ensure that all dyons inside the ``$\rmax$ sphere'' are considered. Since spheres associated with points $\v r$ near the boundary reach into neighboring volumes, one needs to shift the \legosphere{} periodically.


\subsection{Evaluating the long-range part}

Introducing the structure functions $S(\v k) = \sum_{j=1}^{n_D}q_je^{-i\vk\vr_j}$ the long range part can be written according to
\begin{align}
 \Phi^{\text{Long}}(\vr) &= \frac{4\pi}{V}\sum_{\vk\neq0}e^{i\vk\vr}\frac{e^{-\lambda^2\vk^2/2}}{\vk^2}\ 
                S(\vk).
\end{align}
Due to the exponential suppression by $e^{-\lambda^2k^2/2}$, it is sufficient to consider momenta \\ $\v k < k_{\max} \propto 1 / \lambda$. Consequently, one has to sum over $\mathcal O(L^3/\lambda^3)$ different momenta $\v k$ and the cost for computing the structure functions is $\mathcal O(V^2 / \lambda^3)$. Since the structure functions $S(\v k)$ do not depend on $\v r$, they have to be computed only once for a given set of dyon positions.

In a second step the long-range part has to be evaluated at $M\propto V$ points $\v r$ amounting again to a computational cost of $\mathcal O(V^2 / \lambda^3)$.


\subsection{Optimizing the trade-off between the short-range and the long-range part}

The computational cost for the whole algorithm is minimized, when the scaling of the short-range and the long-range part are identical, i.e.\ if $\mathcal O(V \lambda^3) = \mathcal O(V^2 / \lambda^3)$. This can easily be achieved by choosing $\lambda\propto V^{1/6}\propto\sqrt{L}$, resulting in a total computational cost of $\mathcal O(V^{3/2})$. A more detailed discussion, in particular, of how to determine the optimal value for $\lambda$ for a specific dyon ensemble, we refer to \cite{Bruckmann:2011yd}.


\section{\label{sec:numerical_results}Numerical results}

\begin{figure}[t!]
 \begin{minipage}{.49\textwidth}
    \centering
    \includegraphics[width=7cm]{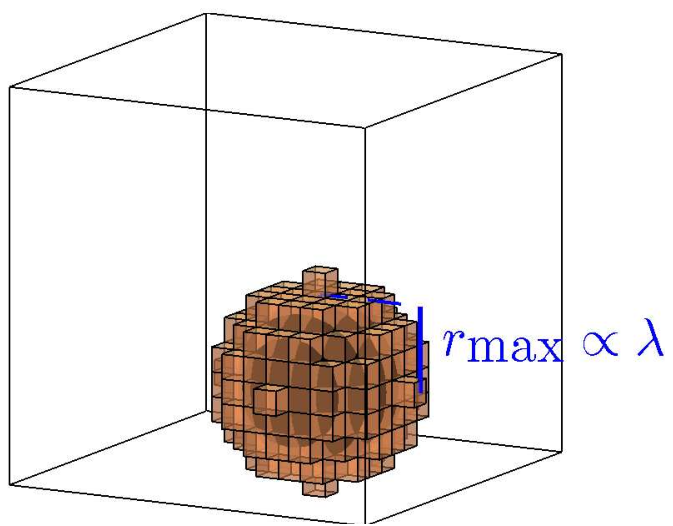}\vspace{0.35cm}
 \caption[\legosphere]{\label{fig:legosphere}The \legosphere{}.}
 \vspace{0.65cm}
 \end{minipage}
 \begin{minipage}{.49\textwidth}
    \centering
    \includegraphics[height=6cm]{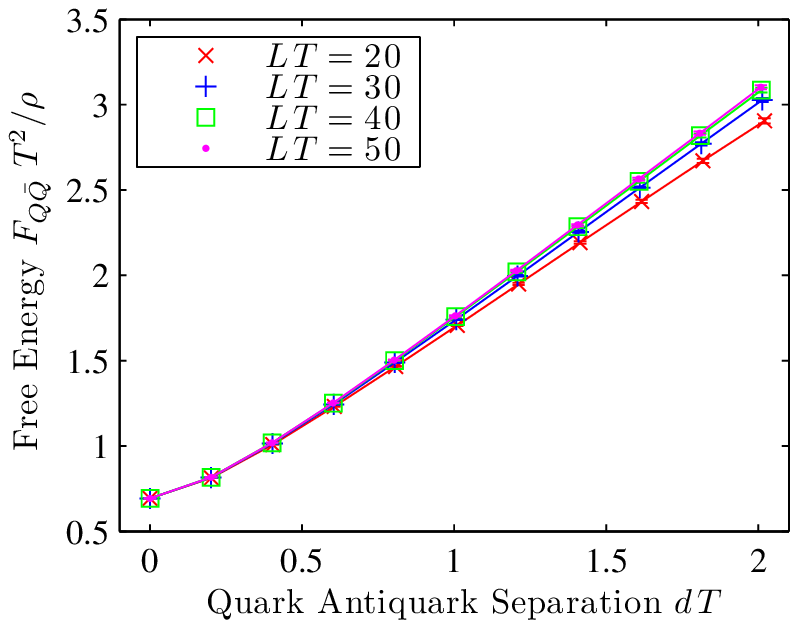}
 \caption{\label{fig:free_energy_comparison}The quark antiquark free energy as a function of the separation $d\,T$ for $T^3/\rho=1$ and finite periodic volume of length $L$.}
 \end{minipage}
\end{figure}

The methods discussed in the previous section were used to evaluate the Polyakov loop correlator for many different dyon ensembles (cf.\ \cite{Bruckmann:2011yd} for a detailed discussion). Selected results for the quark antiquark free energy as a function of the separation in the case of non-interacting dyons are shown in \fig{\ref{fig:free_energy_comparison}}. These results correspond to $\rho/T^3=1$ and four different volumes $L\,T\in\{10,20,30,40\}$, which is equivalent to $n_D\in\{1.000 \, , \, 8.000 \, , \, 27.000 \, , \, 64.000\}$ (we express all dimensionful quantities in units of the temperature $T$). The curves grow linearly at large quark antiquark separations $d\,T$ and seem to converge to an infinite volume curve with increasing $L\,T$.

\begin{figure}[tbp]
 \centering
  \includegraphics[width=14cm]{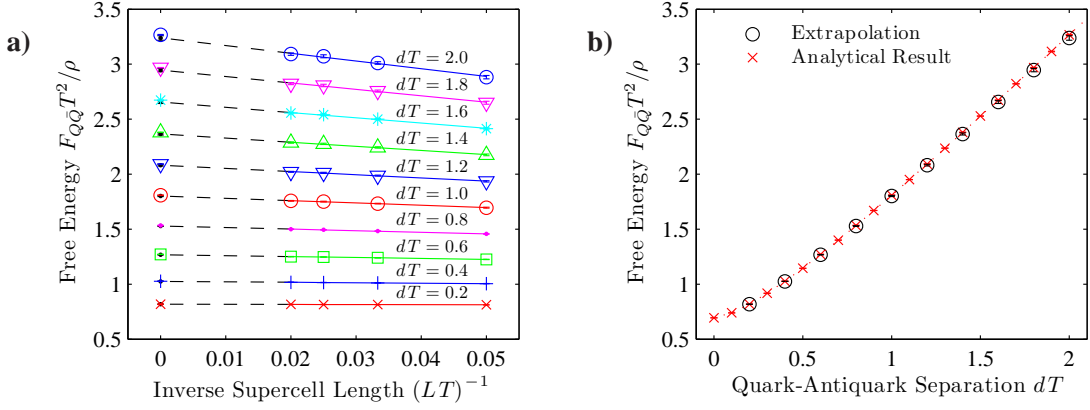}
 \begin{flushleft}
  \vspace{-5.6cm}
    \textbf{a)}
  \hspace{7.2cm}
    \textbf{b)}
  \vspace{4.6cm}
 \end{flushleft}
 \caption{\label{fig:free_energy_extrapolation}\textbf{a)} Quark antiquark free energy as a function of the inverse length of the volume $1/L\,T$ for different separations $d\,T$ and extrapolations to $1/L\,T=0$. The solid curves indicate the fitting range, whereas the dashed lines mark the extrapolations. \textbf{b)} Comparison of the numerical and the analytical infinite volume quark antiquark free energy at these. All data was obtained using $\rho/T^3=1$. }
\end{figure}

To determine this infinite volume curve, we perform a linear $\chi^2$ minimizing fit in $1 / L\,T$ for each quark antiquark separation and extrapolate to infinite volume, corresponding to $1/L\,T = 0$ (cf.\ \fig{\ref{fig:free_energy_extrapolation}a}). The colored points at $1/L\,T = 0$ are analytical results derived in \cite{Bruckmann:2011yd}. The numerical extrapolations are in agreement to these analytical results within statistical errors. The infinite volume free energies (both the numerical Ewald result as well as the analytical result) are shown in \fig{\ref{fig:free_energy_extrapolation}b}, again demonstrating that by means of Ewald's method one can reliably and efficiently determine an infinite volume quark antiquark potential in a long-range dyon ensemble.

\section{Generalization of Ewald's method for arbitrary $1 / r^p$ long-range potentials}

In the previous section we presented numerical results for non-interacting dyons. A more realistic model would be to take dyon interactions into account originating from their moduli space metric. The problem of non-positive-definiteness of the moduli space metric proposed in \cite{Diakonov:2007nv} can be cured by only considering two-dyon, but no three-dyon, four-dyon, etc.\ interactions \cite{Bruckmann:2009nw,Maier:2011}. One then obtains an effective dyon action
\begin{equation}
  \label{eq:action}
  S^{\text{eff}}(\{\v r_k\}) = \frac12\sum_{i=1}^{n_D}\sum^{n_D}_{j=1,j\neq i}\underbrace{\ln\l(1-\frac{2q_iq_j}{\pi T\l|\v r_i - \v r_j\r|}\r)}_{\psi(|\v r_i - \v r_j|)}.
\end{equation}
This effective action also contains ``long-range potentials'' $\psi$, which can be expanded in a power series with respect to inverse dyon separations $r$,
\begin{equation}
 \psi(r) = \frac{\#}{r} +
           \frac{\#}{r^2} +
           \frac{\#}{r^3} + \ldots
\end{equation}
The individual terms $1/r^p$, $p\geq 1$ of this series may be treated with a generalization of Ewald's method to arbitrary powers $1/r^p$ (cf.\ e.g.\ \cite{Essmann:1995}), where the short- and long-range parts are given by
\begin{align}
    \Phi^{\text{Short}}_p(\v r) &= \,\sum_{\v n}\sum_{j=1}^{n_D}\frac{q_j(p)}{|\vr -\vr_j+\v nL|^p}\ g_p\l(\frac{|\vr -\vr_j+\v nL|}{\sqrt{2}\lambda}\r) \\
    \Phi^{\text{Long} }_p(\v r) &= \frac{\pi^{3/2}}{V\l(\sqrt 2\lambda\r)^{p-3}}\sum_{\v k}\sum_{j=1}^{n_D}q_j(p)\,\exp\Big(i\,\v k(\v r-\v r_j)\Big)\ f_p\l(\frac{k\lambda}{\sqrt2}\r),
\end{align}
where the charge $q_j(p)$ of dyon $j$ may depend on the power $p$. Again exponential convergence is guaranteed, due to
\begin{align}
    g_p(x)&=\frac{2}{\Gamma(p/2)}\int_x^{\infty}s^{p-1}\,\exp\l(-s^2\r)\,\d s, \\
    f_p(x)&=\frac{2x^{p-3}}{\Gamma(p/2)}\int_x^{\infty}s^{2-p}\,\exp\l(-s^2\r)\,\d s.
\end{align}


\section{Summary and Outlook}

We applied Ewald's method to simulate non-interacting long-range dyon ensembles numerically. We have demonstrated that this is an efficient method suited to extract the infinite volume quark antiquark free energy. A generalization of Ewald's method offers the possibility to also simulate interacting dyons. Such investigations might help to understand the effects and implications of the moduli space metric of dyons on the quark antiquark free energy and their relevance regarding the phenomenon of confinement. Another interesting aspect would be a possible generalization to manifestly non-Abelian and typically non-rotationally invariant objects like regular gauge instantons, merons, or meron pairs, which also seem to yield confinement (cf.\ e.g.\ \cite{Lenz:2003jp,Wagner:2006qn,Zimmermann:2012zi}). An alternative method to enforce periodic boundary conditions for such non-Abelian long-range objects has already been proposed and tested in \cite{Szasz:2008qk}.


\section*{Acknowledgements}

The authors express their gratitude for financial support by the German Research Foundation (DFG) with various grants: M.W. by the Emmy Noether Programme with grant WA 3000/1-1 and F.B. with grant BR 2872/4-2. This work was supported in part by the Helmholtz International Center for FAIR within the framework of the LOEWE program launched by the State of Hesse.


\end{document}